\documentstyle[12pt,psfig,epsf]{article}
\begin{document}
\begin{flushright}
Preprint SSU-HEP-00/05\\
Samara State University
\end{flushright}

\vspace{30mm}
\immediate\write16{<<WARNING: LINEDRAW macros work with emTeX-dvivers
		    and other drivers supporting emTeX \special's
		    (dviscr, dvihplj, dvidot, dvips, dviwin, etc.) >>}
\newdimen\Lengthunit	   \Lengthunit	= 1.5cm
\newcount\Nhalfperiods	   \Nhalfperiods= 9
\newcount\magnitude	   \magnitude = 1000

\catcode`\*=11
\newdimen\L*   \newdimen\d*   \newdimen\d**
\newdimen\dm*  \newdimen\dd*  \newdimen\dt*
\newdimen\a*   \newdimen\b*   \newdimen\c*
\newdimen\a**  \newdimen\b**
\newdimen\xL*  \newdimen\yL*
\newdimen\rx*  \newdimen\ry*
\newdimen\tmp* \newdimen\linwid*

\newcount\k*   \newcount\l*   \newcount\m*
\newcount\k**  \newcount\l**  \newcount\m**
\newcount\n*   \newcount\dn*  \newcount\r*
\newcount\N*   \newcount\*one \newcount\*two  \*one=1 \*two=2
\newcount\*ths \*ths=1000
\newcount\angle*  \newcount\q*	\newcount\q**
\newcount\angle** \angle**=0
\newcount\sc*	  \sc*=0

\newtoks\cos*  \cos*={1}
\newtoks\sin*  \sin*={0}

\catcode`\[=13

\def\rotate(#1){\advance\angle**#1\angle*=\angle**
\q**=\angle*\ifnum\q**<0\q**=-\q**\fi
\ifnum\q**>360\q*=\angle*\divide\q*360\multiply\q*360\advance\angle*-\q*\fi
\ifnum\angle*<0\advance\angle*360\fi\q**=\angle*\divide\q**90\q**=\q**
\def\sgcos*{+}\def\sgsin*{+}\relax
\ifcase\q**\or
 \def\sgcos*{-}\def\sgsin*{+}\or
 \def\sgcos*{-}\def\sgsin*{-}\or
 \def\sgcos*{+}\def\sgsin*{-}\else\fi
\q*=\q**
\multiply\q*90\advance\angle*-\q*
\ifnum\angle*>45\sc*=1\angle*=-\angle*\advance\angle*90\else\sc*=0\fi
\def[##1,##2]{\ifnum\sc*=0\relax
\edef\cs*{\sgcos*.##1}\edef\sn*{\sgsin*.##2}\ifcase\q**\or
 \edef\cs*{\sgcos*.##2}\edef\sn*{\sgsin*.##1}\or
 \edef\cs*{\sgcos*.##1}\edef\sn*{\sgsin*.##2}\or
 \edef\cs*{\sgcos*.##2}\edef\sn*{\sgsin*.##1}\else\fi\else
\edef\cs*{\sgcos*.##2}\edef\sn*{\sgsin*.##1}\ifcase\q**\or
 \edef\cs*{\sgcos*.##1}\edef\sn*{\sgsin*.##2}\or
 \edef\cs*{\sgcos*.##2}\edef\sn*{\sgsin*.##1}\or
 \edef\cs*{\sgcos*.##1}\edef\sn*{\sgsin*.##2}\else\fi\fi
\cos*={\cs*}\sin*={\sn*}\global\edef\gcos*{\cs*}\global\edef\gsin*{\sn*}}\relax
\ifcase\angle*[9999,0]\or
[999,017]\or[999,034]\or[998,052]\or[997,069]\or[996,087]\or
[994,104]\or[992,121]\or[990,139]\or[987,156]\or[984,173]\or
[981,190]\or[978,207]\or[974,224]\or[970,241]\or[965,258]\or
[961,275]\or[956,292]\or[951,309]\or[945,325]\or[939,342]\or
[933,358]\or[927,374]\or[920,390]\or[913,406]\or[906,422]\or
[898,438]\or[891,453]\or[882,469]\or[874,484]\or[866,499]\or
[857,515]\or[848,529]\or[838,544]\or[829,559]\or[819,573]\or
[809,587]\or[798,601]\or[788,615]\or[777,629]\or[766,642]\or
[754,656]\or[743,669]\or[731,681]\or[719,694]\or[707,707]\or
\else[9999,0]\fi}

\catcode`\[=12

\def\GRAPH(hsize=#1)#2{\hbox to #1\Lengthunit{#2\hss}}

\def\Linewidth#1{\global\linwid*=#1\relax
\global\divide\linwid*10\global\multiply\linwid*\mag
\global\divide\linwid*100\special{em:linewidth \the\linwid*}}

\Linewidth{.4pt}
\def\sm*{\special{em:moveto}}
\def\sl*{\special{em:lineto}}
\let\moveto=\sm*
\let\lineto=\sl*
\newbox\spm*   \newbox\spl*
\setbox\spm*\hbox{\sm*}
\setbox\spl*\hbox{\sl*}

\def\mov#1(#2,#3)#4{\rlap{\L*=#1\Lengthunit
\xL*=#2\L* \yL*=#3\L*
\xL*=\xscale\xL* \yL*=\yscale\yL*
\rx* \the\cos*\xL* \tmp* \the\sin*\yL* \advance\rx*-\tmp*
\ry* \the\cos*\yL* \tmp* \the\sin*\xL* \advance\ry*\tmp*
\kern\rx*\raise\ry*\hbox{#4}}}

\def\rmov*(#1,#2)#3{\rlap{\xL*=#1\yL*=#2\relax
\rx* \the\cos*\xL* \tmp* \the\sin*\yL* \advance\rx*-\tmp*
\ry* \the\cos*\yL* \tmp* \the\sin*\xL* \advance\ry*\tmp*
\kern\rx*\raise\ry*\hbox{#3}}}

\def\lin#1(#2,#3){\rlap{\sm*\mov#1(#2,#3){\sl*}}}

\def\arr*(#1,#2,#3){\rmov*(#1\dd*,#1\dt*){\sm*
\rmov*(#2\dd*,#2\dt*){\rmov*(#3\dt*,-#3\dd*){\sl*}}\sm*
\rmov*(#2\dd*,#2\dt*){\rmov*(-#3\dt*,#3\dd*){\sl*}}}}

\def\arrow#1(#2,#3){\rlap{\lin#1(#2,#3)\mov#1(#2,#3){\relax
\d**=-.012\Lengthunit\dd*=#2\d**\dt*=#3\d**
\arr*(1,10,4)\arr*(3,8,4)\arr*(4.8,4.2,3)}}}

\def\arrlin#1(#2,#3){\rlap{\L*=#1\Lengthunit\L*=.5\L*
\lin#1(#2,#3)\rmov*(#2\L*,#3\L*){\arrow.1(#2,#3)}}}

\def\dasharrow#1(#2,#3){\rlap{{\Lengthunit=0.9\Lengthunit
\dashlin#1(#2,#3)\mov#1(#2,#3){\sm*}}\mov#1(#2,#3){\sl*
\d**=-.012\Lengthunit\dd*=#2\d**\dt*=#3\d**
\arr*(1,10,4)\arr*(3,8,4)\arr*(4.8,4.2,3)}}}

\def\clap#1{\hbox to 0pt{\hss #1\hss}}

\def\ind(#1,#2)#3{\rlap{\L*=.1\Lengthunit
\xL*=#1\L* \yL*=#2\L*
\rx* \the\cos*\xL* \tmp* \the\sin*\yL* \advance\rx*-\tmp*
\ry* \the\cos*\yL* \tmp* \the\sin*\xL* \advance\ry*\tmp*
\kern\rx*\raise\ry*\hbox{\lower2pt\clap{$#3$}}}}

\def\sh*(#1,#2)#3{\rlap{\dm*=\the\n*\d**
\xL*=\xscale\dm* \yL*=\yscale\dm* \xL*=#1\xL* \yL*=#2\yL*
\rx* \the\cos*\xL* \tmp* \the\sin*\yL* \advance\rx*-\tmp*
\ry* \the\cos*\yL* \tmp* \the\sin*\xL* \advance\ry*\tmp*
\kern\rx*\raise\ry*\hbox{#3}}}

\def\calcnum*#1(#2,#3){\a*=1000sp\b*=1000sp\a*=#2\a*\b*=#3\b*
\ifdim\a*<0pt\a*-\a*\fi\ifdim\b*<0pt\b*-\b*\fi
\ifdim\a*>\b*\c*=.96\a*\advance\c*.4\b*
\else\c*=.96\b*\advance\c*.4\a*\fi
\k*\a*\multiply\k*\k*\l*\b*\multiply\l*\l*
\m*\k*\advance\m*\l*\n*\c*\r*\n*\multiply\n*\n*
\dn*\m*\advance\dn*-\n*\divide\dn*2\divide\dn*\r*
\advance\r*\dn*
\c*=\the\Nhalfperiods5sp\c*=#1\c*\ifdim\c*<0pt\c*-\c*\fi
\multiply\c*\r*\N*\c*\divide\N*10000}

\def\dashlin#1(#2,#3){\rlap{\calcnum*#1(#2,#3)\relax
\d**=#1\Lengthunit\ifdim\d**<0pt\d**-\d**\fi
\divide\N*2\multiply\N*2\advance\N*\*one
\divide\d**\N*\sm*\n*\*one\sh*(#2,#3){\sl*}\loop
\advance\n*\*one\sh*(#2,#3){\sm*}\advance\n*\*one
\sh*(#2,#3){\sl*}\ifnum\n*<\N*\repeat}}

\def\dashdotlin#1(#2,#3){\rlap{\calcnum*#1(#2,#3)\relax
\d**=#1\Lengthunit\ifdim\d**<0pt\d**-\d**\fi
\divide\N*2\multiply\N*2\advance\N*1\multiply\N*2\relax
\divide\d**\N*\sm*\n*\*two\sh*(#2,#3){\sl*}\loop
\advance\n*\*one\sh*(#2,#3){\kern-1.48pt\lower.5pt\hbox{\rm.}}\relax
\advance\n*\*one\sh*(#2,#3){\sm*}\advance\n*\*two
\sh*(#2,#3){\sl*}\ifnum\n*<\N*\repeat}}

\def\shl*(#1,#2)#3{\kern#1#3\lower#2#3\hbox{\unhcopy\spl*}}

\def\trianglin#1(#2,#3){\rlap{\toks0={#2}\toks1={#3}\calcnum*#1(#2,#3)\relax
\dd*=.57\Lengthunit\dd*=#1\dd*\divide\dd*\N*
\divide\dd*\*ths \multiply\dd*\magnitude
\d**=#1\Lengthunit\ifdim\d**<0pt\d**-\d**\fi
\multiply\N*2\divide\d**\N*\sm*\n*\*one\loop
\shl**{\dd*}\dd*-\dd*\advance\n*2\relax
\ifnum\n*<\N*\repeat\n*\N*\shl**{0pt}}}

\def\wavelin#1(#2,#3){\rlap{\toks0={#2}\toks1={#3}\calcnum*#1(#2,#3)\relax
\dd*=.23\Lengthunit\dd*=#1\dd*\divide\dd*\N*
\divide\dd*\*ths \multiply\dd*\magnitude
\d**=#1\Lengthunit\ifdim\d**<0pt\d**-\d**\fi
\multiply\N*4\divide\d**\N*\sm*\n*\*one\loop
\shl**{\dd*}\dt*=1.3\dd*\advance\n*\*one
\shl**{\dt*}\advance\n*\*one
\shl**{\dd*}\advance\n*\*two
\dd*-\dd*\ifnum\n*<\N*\repeat\n*\N*\shl**{0pt}}}

\def\w*lin(#1,#2){\rlap{\toks0={#1}\toks1={#2}\d**=\Lengthunit\dd*=-.12\d**
\divide\dd*\*ths \multiply\dd*\magnitude
\N*8\divide\d**\N*\sm*\n*\*one\loop
\shl**{\dd*}\dt*=1.3\dd*\advance\n*\*one
\shl**{\dt*}\advance\n*\*one
\shl**{\dd*}\advance\n*\*one
\shl**{0pt}\dd*-\dd*\advance\n*1\ifnum\n*<\N*\repeat}}

\def\l*arc(#1,#2)[#3][#4]{\rlap{\toks0={#1}\toks1={#2}\d**=\Lengthunit
\dd*=#3.037\d**\dd*=#4\dd*\dt*=#3.049\d**\dt*=#4\dt*\ifdim\d**>10mm\relax
\d**=.25\d**\n*\*one\shl**{-\dd*}\n*\*two\shl**{-\dt*}\n*3\relax
\shl**{-\dd*}\n*4\relax\shl**{0pt}\else
\ifdim\d**>5mm\d**=.5\d**\n*\*one\shl**{-\dt*}\n*\*two
\shl**{0pt}\else\n*\*one\shl**{0pt}\fi\fi}}

\def\d*arc(#1,#2)[#3][#4]{\rlap{\toks0={#1}\toks1={#2}\d**=\Lengthunit
\dd*=#3.037\d**\dd*=#4\dd*\d**=.25\d**\sm*\n*\*one\shl**{-\dd*}\relax
\n*3\relax\sh*(#1,#2){\xL*=\xscale\dd*\yL*=\yscale\dd*
\kern#2\xL*\lower#1\yL*\hbox{\sm*}}\n*4\relax\shl**{0pt}}}

\def\shl**#1{\c*=\the\n*\d**\d*=#1\relax
\a*=\the\toks0\c*\b*=\the\toks1\d*\advance\a*-\b*
\b*=\the\toks1\c*\d*=\the\toks0\d*\advance\b*\d*
\a*=\xscale\a*\b*=\yscale\b*
\rx* \the\cos*\a* \tmp* \the\sin*\b* \advance\rx*-\tmp*
\ry* \the\cos*\b* \tmp* \the\sin*\a* \advance\ry*\tmp*
\raise\ry*\rlap{\kern\rx*\unhcopy\spl*}}

\def\wlin*#1(#2,#3)[#4]{\rlap{\toks0={#2}\toks1={#3}\relax
\c*=#1\l*\c*\c*=.01\Lengthunit\m*\c*\divide\l*\m*
\c*=\the\Nhalfperiods5sp\multiply\c*\l*\N*\c*\divide\N*\*ths
\divide\N*2\multiply\N*2\advance\N*\*one
\dd*=.002\Lengthunit\dd*=#4\dd*\multiply\dd*\l*\divide\dd*\N*
\divide\dd*\*ths \multiply\dd*\magnitude
\d**=#1\multiply\N*4\divide\d**\N*\sm*\n*\*one\loop
\shl**{\dd*}\dt*=1.3\dd*\advance\n*\*one
\shl**{\dt*}\advance\n*\*one
\shl**{\dd*}\advance\n*\*two
\dd*-\dd*\ifnum\n*<\N*\repeat\n*\N*\shl**{0pt}}}

\def\wavebox#1{\setbox0\hbox{#1}\relax
\a*=\wd0\advance\a*14pt\b*=\ht0\advance\b*\dp0\advance\b*14pt\relax
\hbox{\kern9pt\relax
\rmov*(0pt,\ht0){\rmov*(-7pt,7pt){\wlin*\a*(1,0)[+]\wlin*\b*(0,-1)[-]}}\relax
\rmov*(\wd0,-\dp0){\rmov*(7pt,-7pt){\wlin*\a*(-1,0)[+]\wlin*\b*(0,1)[-]}}\relax
\box0\kern9pt}}

\def\rectangle#1(#2,#3){\relax
\lin#1(#2,0)\lin#1(0,#3)\mov#1(0,#3){\lin#1(#2,0)}\mov#1(#2,0){\lin#1(0,#3)}}

\def\dashrectangle#1(#2,#3){\dashlin#1(#2,0)\dashlin#1(0,#3)\relax
\mov#1(0,#3){\dashlin#1(#2,0)}\mov#1(#2,0){\dashlin#1(0,#3)}}

\def\waverectangle#1(#2,#3){\L*=#1\Lengthunit\a*=#2\L*\b*=#3\L*
\ifdim\a*<0pt\a*-\a*\def\x*{-1}\else\def\x*{1}\fi
\ifdim\b*<0pt\b*-\b*\def\y*{-1}\else\def\y*{1}\fi
\wlin*\a*(\x*,0)[-]\wlin*\b*(0,\y*)[+]\relax
\mov#1(0,#3){\wlin*\a*(\x*,0)[+]}\mov#1(#2,0){\wlin*\b*(0,\y*)[-]}}

\def\calcparab*{\ifnum\n*>\m*\k*\N*\advance\k*-\n*\else\k*\n*\fi
\a*=\the\k* sp\a*=10\a*\b*\dm*\advance\b*-\a*\k*\b*
\a*=\the\*ths\b*\divide\a*\l*\multiply\a*\k*
\divide\a*\l*\k*\*ths\r*\a*\advance\k*-\r*\dt*=\the\k*\L*}

\def\arcto#1(#2,#3)[#4]{\rlap{\toks0={#2}\toks1={#3}\calcnum*#1(#2,#3)\relax
\dm*=135sp\dm*=#1\dm*\d**=#1\Lengthunit\ifdim\dm*<0pt\dm*-\dm*\fi
\multiply\dm*\r*\a*=.3\dm*\a*=#4\a*\ifdim\a*<0pt\a*-\a*\fi
\advance\dm*\a*\N*\dm*\divide\N*10000\relax
\divide\N*2\multiply\N*2\advance\N*\*one
\L*=-.25\d**\L*=#4\L*\divide\d**\N*\divide\L*\*ths
\m*\N*\divide\m*2\dm*=\the\m*5sp\l*\dm*\sm*\n*\*one\loop
\calcparab*\shl**{-\dt*}\advance\n*1\ifnum\n*<\N*\repeat}}

\def\arrarcto#1(#2,#3)[#4]{\L*=#1\Lengthunit\L*=.54\L*
\arcto#1(#2,#3)[#4]\rmov*(#2\L*,#3\L*){\d*=.457\L*\d*=#4\d*\d**-\d*
\rmov*(#3\d**,#2\d*){\arrow.02(#2,#3)}}}

\def\dasharcto#1(#2,#3)[#4]{\rlap{\toks0={#2}\toks1={#3}\relax
\calcnum*#1(#2,#3)\dm*=\the\N*5sp\a*=.3\dm*\a*=#4\a*\ifdim\a*<0pt\a*-\a*\fi
\advance\dm*\a*\N*\dm*
\divide\N*20\multiply\N*2\advance\N*1\d**=#1\Lengthunit
\L*=-.25\d**\L*=#4\L*\divide\d**\N*\divide\L*\*ths
\m*\N*\divide\m*2\dm*=\the\m*5sp\l*\dm*
\sm*\n*\*one\loop\calcparab*
\shl**{-\dt*}\advance\n*1\ifnum\n*>\N*\else\calcparab*
\sh*(#2,#3){\xL*=#3\dt* \yL*=#2\dt*
\rx* \the\cos*\xL* \tmp* \the\sin*\yL* \advance\rx*\tmp*
\ry* \the\cos*\yL* \tmp* \the\sin*\xL* \advance\ry*-\tmp*
\kern\rx*\lower\ry*\hbox{\sm*}}\fi
\advance\n*1\ifnum\n*<\N*\repeat}}

\def\*shl*#1{\c*=\the\n*\d**\advance\c*#1\a**\d*\dt*\advance\d*#1\b**
\a*=\the\toks0\c*\b*=\the\toks1\d*\advance\a*-\b*
\b*=\the\toks1\c*\d*=\the\toks0\d*\advance\b*\d*
\rx* \the\cos*\a* \tmp* \the\sin*\b* \advance\rx*-\tmp*
\ry* \the\cos*\b* \tmp* \the\sin*\a* \advance\ry*\tmp*
\raise\ry*\rlap{\kern\rx*\unhcopy\spl*}}

\def\calcnormal*#1{\b**=10000sp\a**\b**\k*\n*\advance\k*-\m*
\multiply\a**\k*\divide\a**\m*\a**=#1\a**\ifdim\a**<0pt\a**-\a**\fi
\ifdim\a**>\b**\d*=.96\a**\advance\d*.4\b**
\else\d*=.96\b**\advance\d*.4\a**\fi
\d*=.01\d*\r*\d*\divide\a**\r*\divide\b**\r*
\ifnum\k*<0\a**-\a**\fi\d*=#1\d*\ifdim\d*<0pt\b**-\b**\fi
\k*\a**\a**=\the\k*\dd*\k*\b**\b**=\the\k*\dd*}

\def\wavearcto#1(#2,#3)[#4]{\rlap{\toks0={#2}\toks1={#3}\relax
\calcnum*#1(#2,#3)\c*=\the\N*5sp\a*=.4\c*\a*=#4\a*\ifdim\a*<0pt\a*-\a*\fi
\advance\c*\a*\N*\c*\divide\N*20\multiply\N*2\advance\N*-1\multiply\N*4\relax
\d**=#1\Lengthunit\dd*=.012\d**
\divide\dd*\*ths \multiply\dd*\magnitude
\ifdim\d**<0pt\d**-\d**\fi\L*=.25\d**
\divide\d**\N*\divide\dd*\N*\L*=#4\L*\divide\L*\*ths
\m*\N*\divide\m*2\dm*=\the\m*0sp\l*\dm*
\sm*\n*\*one\loop\calcnormal*{#4}\calcparab*
\*shl*{1}\advance\n*\*one\calcparab*
\*shl*{1.3}\advance\n*\*one\calcparab*
\*shl*{1}\advance\n*2\dd*-\dd*\ifnum\n*<\N*\repeat\n*\N*\shl**{0pt}}}

\def\triangarcto#1(#2,#3)[#4]{\rlap{\toks0={#2}\toks1={#3}\relax
\calcnum*#1(#2,#3)\c*=\the\N*5sp\a*=.4\c*\a*=#4\a*\ifdim\a*<0pt\a*-\a*\fi
\advance\c*\a*\N*\c*\divide\N*20\multiply\N*2\advance\N*-1\multiply\N*2\relax
\d**=#1\Lengthunit\dd*=.012\d**
\divide\dd*\*ths \multiply\dd*\magnitude
\ifdim\d**<0pt\d**-\d**\fi\L*=.25\d**
\divide\d**\N*\divide\dd*\N*\L*=#4\L*\divide\L*\*ths
\m*\N*\divide\m*2\dm*=\the\m*0sp\l*\dm*
\sm*\n*\*one\loop\calcnormal*{#4}\calcparab*
\*shl*{1}\advance\n*2\dd*-\dd*\ifnum\n*<\N*\repeat\n*\N*\shl**{0pt}}}

\def\hr*#1{\L*=\xscale\Lengthunit\ifnum
\angle**=0\clap{\vrule width#1\L* height.1pt}\else
\L*=#1\L*\L*=.5\L*\rmov*(-\L*,0pt){\sm*}\rmov*(\L*,0pt){\sl*}\fi}

\def\shade#1[#2]{\rlap{\Lengthunit=#1\Lengthunit
\special{em:linewidth .001pt}\relax
\mov(0,#2.05){\hr*{.994}}\mov(0,#2.1){\hr*{.980}}\relax
\mov(0,#2.15){\hr*{.953}}\mov(0,#2.2){\hr*{.916}}\relax
\mov(0,#2.25){\hr*{.867}}\mov(0,#2.3){\hr*{.798}}\relax
\mov(0,#2.35){\hr*{.715}}\mov(0,#2.4){\hr*{.603}}\relax
\mov(0,#2.45){\hr*{.435}}\special{em:linewidth \the\linwid*}}}

\def\dshade#1[#2]{\rlap{\special{em:linewidth .001pt}\relax
\Lengthunit=#1\Lengthunit\if#2-\def\t*{+}\else\def\t*{-}\fi
\mov(0,\t*.025){\relax
\mov(0,#2.05){\hr*{.995}}\mov(0,#2.1){\hr*{.988}}\relax
\mov(0,#2.15){\hr*{.969}}\mov(0,#2.2){\hr*{.937}}\relax
\mov(0,#2.25){\hr*{.893}}\mov(0,#2.3){\hr*{.836}}\relax
\mov(0,#2.35){\hr*{.760}}\mov(0,#2.4){\hr*{.662}}\relax
\mov(0,#2.45){\hr*{.531}}\mov(0,#2.5){\hr*{.320}}\relax
\special{em:linewidth \the\linwid*}}}}

\def\vdot{\rlap{\kern-1.9pt\lower1.8pt\hbox{$\scriptstyle\bullet$}}}
\def\vtimes{\rlap{\kern-3pt\lower1.8pt\hbox{$\scriptstyle\times$}}}
\def\vDot{\rlap{\kern-2.3pt\lower2.7pt\hbox{$\bullet$}}}
\def\vTimes{\rlap{\kern-3.6pt\lower2.4pt\hbox{$\times$}}}

\def\arc(#1)[#2,#3]{{\k*=#2\l*=#3\m*=\l*
\advance\m*-6\ifnum\k*>\l*\relax\else
{\rotate(#2)\mov(#1,0){\sm*}}\loop
\ifnum\k*<\m*\advance\k*5{\rotate(\k*)\mov(#1,0){\sl*}}\repeat
{\rotate(#3)\mov(#1,0){\sl*}}\fi}}

\def\dasharc(#1)[#2,#3]{{\k**=#2\n*=#3\advance\n*-1\advance\n*-\k**
\L*=1000sp\L*#1\L* \multiply\L*\n* \multiply\L*\Nhalfperiods
\divide\L*57\N*\L* \divide\N*2000\ifnum\N*=0\N*1\fi
\r*\n*	\divide\r*\N* \ifnum\r*<2\r*2\fi
\m**\r* \divide\m**2 \l**\r* \advance\l**-\m** \N*\n* \divide\N*\r*
\k**\r* \multiply\k**\N* \dn*\n* \advance\dn*-\k** \divide\dn*2\advance\dn*\*one
\r*\l** \divide\r*2\advance\dn*\r* \advance\N*-2\k**#2\relax
\ifnum\l**<6{\rotate(#2)\mov(#1,0){\sm*}}\advance\k**\dn*
{\rotate(\k**)\mov(#1,0){\sl*}}\advance\k**\m**
{\rotate(\k**)\mov(#1,0){\sm*}}\loop
\advance\k**\l**{\rotate(\k**)\mov(#1,0){\sl*}}\advance\k**\m**
{\rotate(\k**)\mov(#1,0){\sm*}}\advance\N*-1\ifnum\N*>0\repeat
{\rotate(#3)\mov(#1,0){\sl*}}\else\advance\k**\dn*
\arc(#1)[#2,\k**]\loop\advance\k**\m** \r*\k**
\advance\k**\l** {\arc(#1)[\r*,\k**]}\relax
\advance\N*-1\ifnum\N*>0\repeat
\advance\k**\m**\arc(#1)[\k**,#3]\fi}}

\def\triangarc#1(#2)[#3,#4]{{\k**=#3\n*=#4\advance\n*-\k**
\L*=1000sp\L*#2\L* \multiply\L*\n* \multiply\L*\Nhalfperiods
\divide\L*57\N*\L* \divide\N*1000\ifnum\N*=0\N*1\fi
\d**=#2\Lengthunit \d*\d** \divide\d*57\multiply\d*\n*
\r*\n*	\divide\r*\N* \ifnum\r*<2\r*2\fi
\m**\r* \divide\m**2 \l**\r* \advance\l**-\m** \N*\n* \divide\N*\r*
\dt*\d* \divide\dt*\N* \dt*.5\dt* \dt*#1\dt*
\divide\dt*1000\multiply\dt*\magnitude
\k**\r* \multiply\k**\N* \dn*\n* \advance\dn*-\k** \divide\dn*2\relax
\r*\l** \divide\r*2\advance\dn*\r* \advance\N*-1\k**#3\relax
{\rotate(#3)\mov(#2,0){\sm*}}\advance\k**\dn*
{\rotate(\k**)\mov(#2,0){\sl*}}\advance\k**-\m**\advance\l**\m**\loop\dt*-\dt*
\d*\d** \advance\d*\dt*
\advance\k**\l**{\rotate(\k**)\rmov*(\d*,0pt){\sl*}}%
\advance\N*-1\ifnum\N*>0\repeat\advance\k**\m**
{\rotate(\k**)\mov(#2,0){\sl*}}{\rotate(#4)\mov(#2,0){\sl*}}}}

\def\wavearc#1(#2)[#3,#4]{{\k**=#3\n*=#4\advance\n*-\k**
\L*=4000sp\L*#2\L* \multiply\L*\n* \multiply\L*\Nhalfperiods
\divide\L*57\N*\L* \divide\N*1000\ifnum\N*=0\N*1\fi
\d**=#2\Lengthunit \d*\d** \divide\d*57\multiply\d*\n*
\r*\n*	\divide\r*\N* \ifnum\r*=0\r*1\fi
\m**\r* \divide\m**2 \l**\r* \advance\l**-\m** \N*\n* \divide\N*\r*
\dt*\d* \divide\dt*\N* \dt*.7\dt* \dt*#1\dt*
\divide\dt*1000\multiply\dt*\magnitude
\k**\r* \multiply\k**\N* \dn*\n* \advance\dn*-\k** \divide\dn*2\relax
\divide\N*4\advance\N*-1\k**#3\relax
{\rotate(#3)\mov(#2,0){\sm*}}\advance\k**\dn*
{\rotate(\k**)\mov(#2,0){\sl*}}\advance\k**-\m**\advance\l**\m**\loop\dt*-\dt*
\d*\d** \advance\d*\dt* \dd*\d** \advance\dd*1.3\dt*
\advance\k**\r*{\rotate(\k**)\rmov*(\d*,0pt){\sl*}}\relax
\advance\k**\r*{\rotate(\k**)\rmov*(\dd*,0pt){\sl*}}\relax
\advance\k**\r*{\rotate(\k**)\rmov*(\d*,0pt){\sl*}}\relax
\advance\k**\r*
\advance\N*-1\ifnum\N*>0\repeat\advance\k**\m**
{\rotate(\k**)\mov(#2,0){\sl*}}{\rotate(#4)\mov(#2,0){\sl*}}}}

\def\gmov*#1(#2,#3)#4{\rlap{\L*=#1\Lengthunit
\xL*=#2\L* \yL*=#3\L*
\rx* \gcos*\xL* \tmp* \gsin*\yL* \advance\rx*-\tmp*
\ry* \gcos*\yL* \tmp* \gsin*\xL* \advance\ry*\tmp*
\rx*=\xscale\rx* \ry*=\yscale\ry*
\xL* \the\cos*\rx* \tmp* \the\sin*\ry* \advance\xL*-\tmp*
\yL* \the\cos*\ry* \tmp* \the\sin*\rx* \advance\yL*\tmp*
\kern\xL*\raise\yL*\hbox{#4}}}

\def\rgmov*(#1,#2)#3{\rlap{\xL*#1\yL*#2\relax
\rx* \gcos*\xL* \tmp* \gsin*\yL* \advance\rx*-\tmp*
\ry* \gcos*\yL* \tmp* \gsin*\xL* \advance\ry*\tmp*
\rx*=\xscale\rx* \ry*=\yscale\ry*
\xL* \the\cos*\rx* \tmp* \the\sin*\ry* \advance\xL*-\tmp*
\yL* \the\cos*\ry* \tmp* \the\sin*\rx* \advance\yL*\tmp*
\kern\xL*\raise\yL*\hbox{#3}}}

\def\Earc(#1)[#2,#3][#4,#5]{{\k*=#2\l*=#3\m*=\l*
\advance\m*-6\ifnum\k*>\l*\relax\else\def\xscale{#4}\def\yscale{#5}\relax
{\angle**0\rotate(#2)}\gmov*(#1,0){\sm*}\loop
\ifnum\k*<\m*\advance\k*5\relax
{\angle**0\rotate(\k*)}\gmov*(#1,0){\sl*}\repeat
{\angle**0\rotate(#3)}\gmov*(#1,0){\sl*}\relax
\def\xscale{1}\def\yscale{1}\fi}}

\def\dashEarc(#1)[#2,#3][#4,#5]{{\k**=#2\n*=#3\advance\n*-1\advance\n*-\k**
\L*=1000sp\L*#1\L* \multiply\L*\n* \multiply\L*\Nhalfperiods
\divide\L*57\N*\L* \divide\N*2000\ifnum\N*=0\N*1\fi
\r*\n*	\divide\r*\N* \ifnum\r*<2\r*2\fi
\m**\r* \divide\m**2 \l**\r* \advance\l**-\m** \N*\n* \divide\N*\r*
\k**\r*\multiply\k**\N* \dn*\n* \advance\dn*-\k** \divide\dn*2\advance\dn*\*one
\r*\l** \divide\r*2\advance\dn*\r* \advance\N*-2\k**#2\relax
\ifnum\l**<6\def\xscale{#4}\def\yscale{#5}\relax
{\angle**0\rotate(#2)}\gmov*(#1,0){\sm*}\advance\k**\dn*
{\angle**0\rotate(\k**)}\gmov*(#1,0){\sl*}\advance\k**\m**
{\angle**0\rotate(\k**)}\gmov*(#1,0){\sm*}\loop
\advance\k**\l**{\angle**0\rotate(\k**)}\gmov*(#1,0){\sl*}\advance\k**\m**
{\angle**0\rotate(\k**)}\gmov*(#1,0){\sm*}\advance\N*-1\ifnum\N*>0\repeat
{\angle**0\rotate(#3)}\gmov*(#1,0){\sl*}\def\xscale{1}\def\yscale{1}\else
\advance\k**\dn* \Earc(#1)[#2,\k**][#4,#5]\loop\advance\k**\m** \r*\k**
\advance\k**\l** {\Earc(#1)[\r*,\k**][#4,#5]}\relax
\advance\N*-1\ifnum\N*>0\repeat
\advance\k**\m**\Earc(#1)[\k**,#3][#4,#5]\fi}}

\def\triangEarc#1(#2)[#3,#4][#5,#6]{{\k**=#3\n*=#4\advance\n*-\k**
\L*=1000sp\L*#2\L* \multiply\L*\n* \multiply\L*\Nhalfperiods
\divide\L*57\N*\L* \divide\N*1000\ifnum\N*=0\N*1\fi
\d**=#2\Lengthunit \d*\d** \divide\d*57\multiply\d*\n*
\r*\n*	\divide\r*\N* \ifnum\r*<2\r*2\fi
\m**\r* \divide\m**2 \l**\r* \advance\l**-\m** \N*\n* \divide\N*\r*
\dt*\d* \divide\dt*\N* \dt*.5\dt* \dt*#1\dt*
\divide\dt*1000\multiply\dt*\magnitude
\k**\r* \multiply\k**\N* \dn*\n* \advance\dn*-\k** \divide\dn*2\relax
\r*\l** \divide\r*2\advance\dn*\r* \advance\N*-1\k**#3\relax
\def\xscale{#5}\def\yscale{#6}\relax
{\angle**0\rotate(#3)}\gmov*(#2,0){\sm*}\advance\k**\dn*
{\angle**0\rotate(\k**)}\gmov*(#2,0){\sl*}\advance\k**-\m**
\advance\l**\m**\loop\dt*-\dt* \d*\d** \advance\d*\dt*
\advance\k**\l**{\angle**0\rotate(\k**)}\rgmov*(\d*,0pt){\sl*}\relax
\advance\N*-1\ifnum\N*>0\repeat\advance\k**\m**
{\angle**0\rotate(\k**)}\gmov*(#2,0){\sl*}\relax
{\angle**0\rotate(#4)}\gmov*(#2,0){\sl*}\def\xscale{1}\def\yscale{1}}}

\def\waveEarc#1(#2)[#3,#4][#5,#6]{{\k**=#3\n*=#4\advance\n*-\k**
\L*=4000sp\L*#2\L* \multiply\L*\n* \multiply\L*\Nhalfperiods
\divide\L*57\N*\L* \divide\N*1000\ifnum\N*=0\N*1\fi
\d**=#2\Lengthunit \d*\d** \divide\d*57\multiply\d*\n*
\r*\n*	\divide\r*\N* \ifnum\r*=0\r*1\fi
\m**\r* \divide\m**2 \l**\r* \advance\l**-\m** \N*\n* \divide\N*\r*
\dt*\d* \divide\dt*\N* \dt*.7\dt* \dt*#1\dt*
\divide\dt*1000\multiply\dt*\magnitude
\k**\r* \multiply\k**\N* \dn*\n* \advance\dn*-\k** \divide\dn*2\relax
\divide\N*4\advance\N*-1\k**#3\def\xscale{#5}\def\yscale{#6}\relax
{\angle**0\rotate(#3)}\gmov*(#2,0){\sm*}\advance\k**\dn*
{\angle**0\rotate(\k**)}\gmov*(#2,0){\sl*}\advance\k**-\m**
\advance\l**\m**\loop\dt*-\dt*
\d*\d** \advance\d*\dt* \dd*\d** \advance\dd*1.3\dt*
\advance\k**\r*{\angle**0\rotate(\k**)}\rgmov*(\d*,0pt){\sl*}\relax
\advance\k**\r*{\angle**0\rotate(\k**)}\rgmov*(\dd*,0pt){\sl*}\relax
\advance\k**\r*{\angle**0\rotate(\k**)}\rgmov*(\d*,0pt){\sl*}\relax
\advance\k**\r*
\advance\N*-1\ifnum\N*>0\repeat\advance\k**\m**
{\angle**0\rotate(\k**)}\gmov*(#2,0){\sl*}\relax
{\angle**0\rotate(#4)}\gmov*(#2,0){\sl*}\def\xscale{1}\def\yscale{1}}}

\newcount\CatcodeOfAtSign
\CatcodeOfAtSign=\the\catcode`\@
\catcode`\@=11
\def\@arc#1[#2][#3]{\rlap{\Lengthunit=#1\Lengthunit
\sm*\l*arc(#2.1914,#3.0381)[#2][#3]\relax
\mov(#2.1914,#3.0381){\l*arc(#2.1622,#3.1084)[#2][#3]}\relax
\mov(#2.3536,#3.1465){\l*arc(#2.1084,#3.1622)[#2][#3]}\relax
\mov(#2.4619,#3.3086){\l*arc(#2.0381,#3.1914)[#2][#3]}}}

\def\dash@arc#1[#2][#3]{\rlap{\Lengthunit=#1\Lengthunit
\d*arc(#2.1914,#3.0381)[#2][#3]\relax
\mov(#2.1914,#3.0381){\d*arc(#2.1622,#3.1084)[#2][#3]}\relax
\mov(#2.3536,#3.1465){\d*arc(#2.1084,#3.1622)[#2][#3]}\relax
\mov(#2.4619,#3.3086){\d*arc(#2.0381,#3.1914)[#2][#3]}}}

\def\wave@arc#1[#2][#3]{\rlap{\Lengthunit=#1\Lengthunit
\w*lin(#2.1914,#3.0381)\relax
\mov(#2.1914,#3.0381){\w*lin(#2.1622,#3.1084)}\relax
\mov(#2.3536,#3.1465){\w*lin(#2.1084,#3.1622)}\relax
\mov(#2.4619,#3.3086){\w*lin(#2.0381,#3.1914)}}}

\def\bezier#1(#2,#3)(#4,#5)(#6,#7){\N*#1\l*\N* \advance\l*\*one
\d* #4\Lengthunit \advance\d* -#2\Lengthunit \multiply\d* \*two
\b* #6\Lengthunit \advance\b* -#2\Lengthunit
\advance\b*-\d* \divide\b*\N*
\d** #5\Lengthunit \advance\d** -#3\Lengthunit \multiply\d** \*two
\b** #7\Lengthunit \advance\b** -#3\Lengthunit
\advance\b** -\d** \divide\b**\N*
\mov(#2,#3){\sm*{\loop\ifnum\m*<\l*
\a*\m*\b* \advance\a*\d* \divide\a*\N* \multiply\a*\m*
\a**\m*\b** \advance\a**\d** \divide\a**\N* \multiply\a**\m*
\rmov*(\a*,\a**){\unhcopy\spl*}\advance\m*\*one\repeat}}}

\catcode`\*=12

\newcount\n@ast
\def\n@ast@#1{\n@ast0\relax\get@ast@#1\end}
\def\get@ast@#1{\ifx#1\end\let\next\relax\else
\ifx#1*\advance\n@ast1\fi\let\next\get@ast@\fi\next}

\newif\if@up \newif\if@dwn
\def\up@down@#1{\@upfalse\@dwnfalse
\if#1u\@uptrue\fi\if#1U\@uptrue\fi\if#1+\@uptrue\fi
\if#1d\@dwntrue\fi\if#1D\@dwntrue\fi\if#1-\@dwntrue\fi}

\def\halfcirc#1(#2)[#3]{{\Lengthunit=#2\Lengthunit\up@down@{#3}\relax
\if@up\mov(0,.5){\@arc[-][-]\@arc[+][-]}\fi
\if@dwn\mov(0,-.5){\@arc[-][+]\@arc[+][+]}\fi
\def\lft{\mov(0,.5){\@arc[-][-]}\mov(0,-.5){\@arc[-][+]}}\relax
\def\rght{\mov(0,.5){\@arc[+][-]}\mov(0,-.5){\@arc[+][+]}}\relax
\if#3l\lft\fi\if#3L\lft\fi\if#3r\rght\fi\if#3R\rght\fi
\n@ast@{#1}\relax
\ifnum\n@ast>0\if@up\shade[+]\fi\if@dwn\shade[-]\fi\fi
\ifnum\n@ast>1\if@up\dshade[+]\fi\if@dwn\dshade[-]\fi\fi}}

\def\halfdashcirc(#1)[#2]{{\Lengthunit=#1\Lengthunit\up@down@{#2}\relax
\if@up\mov(0,.5){\dash@arc[-][-]\dash@arc[+][-]}\fi
\if@dwn\mov(0,-.5){\dash@arc[-][+]\dash@arc[+][+]}\fi
\def\lft{\mov(0,.5){\dash@arc[-][-]}\mov(0,-.5){\dash@arc[-][+]}}\relax
\def\rght{\mov(0,.5){\dash@arc[+][-]}\mov(0,-.5){\dash@arc[+][+]}}\relax
\if#2l\lft\fi\if#2L\lft\fi\if#2r\rght\fi\if#2R\rght\fi}}

\def\halfwavecirc(#1)[#2]{{\Lengthunit=#1\Lengthunit\up@down@{#2}\relax
\if@up\mov(0,.5){\wave@arc[-][-]\wave@arc[+][-]}\fi
\if@dwn\mov(0,-.5){\wave@arc[-][+]\wave@arc[+][+]}\fi
\def\lft{\mov(0,.5){\wave@arc[-][-]}\mov(0,-.5){\wave@arc[-][+]}}\relax
\def\rght{\mov(0,.5){\wave@arc[+][-]}\mov(0,-.5){\wave@arc[+][+]}}\relax
\if#2l\lft\fi\if#2L\lft\fi\if#2r\rght\fi\if#2R\rght\fi}}

\catcode`\*=11

\def\Circle#1(#2){\halfcirc#1(#2)[u]\halfcirc#1(#2)[d]\n@ast@{#1}\relax
\ifnum\n@ast>0\L*=\xscale\Lengthunit
\ifnum\angle**=0\clap{\vrule width#2\L* height.1pt}\else
\L*=#2\L*\L*=.5\L*\special{em:linewidth .001pt}\relax
\rmov*(-\L*,0pt){\sm*}\rmov*(\L*,0pt){\sl*}\relax
\special{em:linewidth \the\linwid*}\fi\fi}

\catcode`\*=12

\def\wavecirc(#1){\halfwavecirc(#1)[u]\halfwavecirc(#1)[d]}

\def\dashcirc(#1){\halfdashcirc(#1)[u]\halfdashcirc(#1)[d]}

\def\xscale{1}
\def\yscale{1}

\def\Ellipse#1(#2)[#3,#4]{\def\xscale{#3}\def\yscale{#4}\relax
\Circle#1(#2)\def\xscale{1}\def\yscale{1}}

\def\dashEllipse(#1)[#2,#3]{\def\xscale{#2}\def\yscale{#3}\relax
\dashcirc(#1)\def\xscale{1}\def\yscale{1}}

\def\waveEllipse(#1)[#2,#3]{\def\xscale{#2}\def\yscale{#3}\relax
\wavecirc(#1)\def\xscale{1}\def\yscale{1}}

\def\halfEllipse#1(#2)[#3][#4,#5]{\def\xscale{#4}\def\yscale{#5}\relax
\halfcirc#1(#2)[#3]\def\xscale{1}\def\yscale{1}}

\def\halfdashEllipse(#1)[#2][#3,#4]{\def\xscale{#3}\def\yscale{#4}\relax
\halfdashcirc(#1)[#2]\def\xscale{1}\def\yscale{1}}

\def\halfwaveEllipse(#1)[#2][#3,#4]{\def\xscale{#3}\def\yscale{#4}\relax
\halfwavecirc(#1)[#2]\def\xscale{1}\def\yscale{1}}

\catcode`\@=\the\CatcodeOfAtSign

\begin{center}
{\bf PROTON POLARIZABILITY CONTRIBUTION\\
TO THE HYDROGEN HYPERFINE SPLITTING}\\

\vspace{4mm}

R.N.~Faustov \\Scientific Council "Cybernetics" RAS\\
117333, Moscow, Vavilov, 40, Russia,\\
A.P.~Martynenko\\ Department of Theoretical Physics, Samara State University,\\
443011, Samara, Pavlov, 1, Russia
\end{center}

\begin{abstract}
The contribution of the proton polarizability to the hydrogen hyperfine
splitting is evaluated on the basis of modern experimental and
theoretical results on the proton polarized structure functions.
The value of this correction is equal to 1.4 {\rm ppm}.
\end{abstract}

\newpage

The investigation of the hyperfine splitting (HFS) of the hydrogen atom ground
state is considered during many years as an important test of quantum
electrodynamics. The experimental value of the hydrogen hyperfine splitting
was obtained with very high accuracy \cite{Hellwig}:
\begin{equation}
\Delta E^{exp}_{HFS}=1420405.7517667(9)~~{\rm kHz}.
\end{equation}

The corresponding theoretical expression of the hydrogen hyperfine
splitting may be written at present time in the form \cite{DTF}:
\begin{equation}
\Delta E^{th}_{HFS}=E_F(1+\delta^{QED}+\delta^S+\delta^P),~~E_F=\frac{8}{3}
\alpha^4\frac{\mu_Pm_p^2m_e^2}{(m_p+m_e)^3},
\end{equation}
where $\mu_P$ is the proton magnetic moment, $m_e$, $m_p$ are the masses
of the electron and proton. The calculation of different corrections to
$E_F$ has a long history. Modern status in the theory of hydrogenic atoms
was presented in details in \cite{EGS}. $\delta^{QED}$ denotes the contribution
of higher-order quantumelectrodynamical effects. Corrections $\delta^S$ and
$\delta^P$ take into account the influence of strong interaction. $\delta^S$
describes the effects of proton finite-size and recoil contribution.
$\delta^P$ is the correction due to the proton polarizability. Basic uncertainty
of theoretical result (2) is related with this term.

The main contribution to $\delta^P$ is determined by two-photon diagrams,
shown in Figure 1. The corresponding amplitudes of virtual Compton
scattering on proton can be expressed through nucleon polarized structure
functions $\rm G_1(\nu,Q^2)$ and $\rm G_2(\nu,Q^2)$. Inelastic contribution
of the diagrams (a), (b) Figure 1 may be presented in the form [3-8]:
\begin{equation}
\Delta E_{HFS}^P=\frac{Z\alpha m_e}{2\pi m_p(1+\kappa)}E_F(\Delta_1+\Delta_2)=
(\delta_1^P+\delta_2^P)E_F=\delta^PE_F,
\end{equation}
\begin{equation}
\Delta_1=\int_0^\infty\frac{dQ^2}{Q^2}\left\{\frac{9}{4}F_2^2(Q^2)-
4m_p^3\int_{\nu_{th}}^\infty\frac{d\nu}{\nu}\beta_1\left(\frac{\nu^2}{Q^2}\right)
G_1(\nu,Q^2)\right\},
\end{equation}
\begin{equation}
\Delta_2=-12m_p^2\int_0^\infty\frac{dQ^2}{Q^2}
\int_{\nu_{th}}^\infty d\nu\beta_2\left(\frac{\nu^2}{Q^2}\right)
G_2(\nu,Q^2),
\end{equation}
where $\nu_{th}$ determines the pion-nucleon threshold:
\begin{equation}
\nu_{th}=m_\pi+\frac{m_\pi^2+Q^2}{2m_p},
\end{equation}
and the functions $\beta_{1,2}$ have the form:
\begin{equation}
\beta_1(\theta)=3\theta-2\theta^2-2(2-\theta)\sqrt{\theta(\theta+1)},
\end{equation}
\begin{equation}
\beta_2(\theta)=1+2\theta-2\sqrt{\theta(\theta+1)},~~\theta=\nu^2/Q^2.
\end{equation}
$F_2(Q^2)$ is the Pauli form factor of the proton, $\kappa$ is the proton
anomalous magnetic moment: $\kappa$=1.792847386(63) \cite{RPP}. During many
years there was not enough experimental data and theoretical information
about proton spin-dependent structure functions. So, the previous study of
the contribution $\Delta E_{HFS}^P$ contains only estimation of the proton
polarizability effects: $\delta^P\sim 1\div 2$ {\rm ppm} or the calculation of main
resonance contributions \cite{V,G,Z,FMS}. The theoretical bound for the proton
polarizability contribution is $|\delta^P|\le 4~{\rm ppm}$. As noted in \cite{EGS},
the problem of the proton polarizability contribution requires new investigation,
which takes into account more recent experimental data on the spin structure
of the nucleon.

The polarized structure functions $g_1(\nu,Q^2)$ and $g_2(\nu,Q^2)$ enter
in the antisymmetric part of hadronic tensor $W_{\mu\nu}$, describing
lepton-nucleon deep inelastic scattering \cite{Close}:
\begin{equation}
W_{\mu\nu}=W_{\mu\nu}^{[S]}+W_{\mu\nu}^{[A]},
\end{equation}
\begin{equation}
W_{\mu\nu}^{[S]}=\left(-g_{\mu\nu}+\frac{q_\mu q_\nu}{q^2}\right)W_1(\nu,Q^2)+
\left(P_\mu-\frac{P\cdot q}{q^2}q_\mu\right)\left(P_\nu-\frac{P\cdot q}{q^2}q_\nu\right)
\frac{W_2(\nu,Q^2)}{m_p^2},
\end{equation}
\begin{equation}
W_{\mu\nu}^{[A]}=\epsilon_{\mu\nu\alpha\beta}q^\alpha\left\{S^\beta
\frac{g_1(\nu,Q^2)}{P\cdot q}+[(P\cdot q)S^\beta-(S\cdot q)P^\beta]
\frac{g_2(\nu,Q^2)}{(P\cdot q)^2}\right\},
\end{equation}
where $g_1(\nu,Q^2)=m_p^2\nu G_1(\nu,Q^2)$,
$g_2(\nu,Q^2)=m_p\nu^2 G_2(\nu,Q^2)$,
$q^2=-Q^2$ is the square of the four-momentum transfer. The invariant
quantity $P\cdot q$ is related to the energy transfer $\nu$ in the proton
rest frame: $P\cdot q=m_p\nu$. The invariant mass of the electroproduced
hadronic system, W, is then $W^2=m_p^2+2m_p\nu-Q^2$.

\begin{figure}
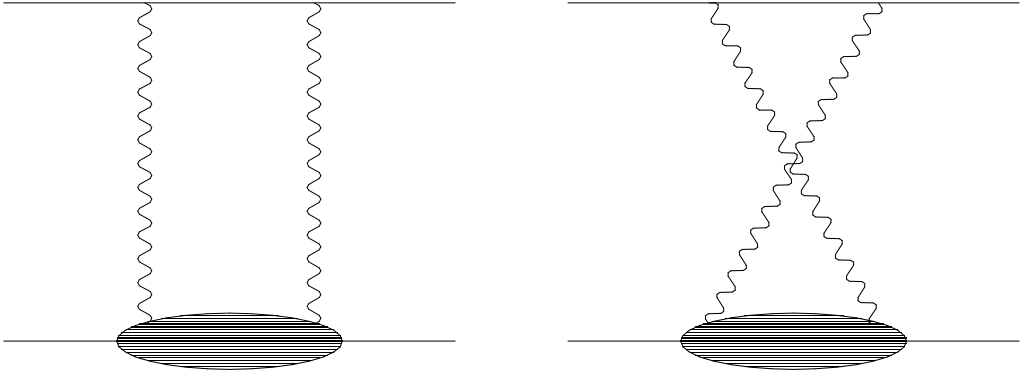

\magnitude=2000
\GRAPH(hsize=15){
\mov(0,0){\lin(1,0)}%
\mov(3,0){\lin(1,0)}%
\mov(5,0){\lin(1,0)}%
\mov(8,0){\lin(1,0)}%
\mov(0,3){\lin(4,0)}%
\mov(5,3){\lin(4,0)}%
\mov(1.25,3){\wavelin(0,-2.85)}%
\mov(2.75,3){\wavelin(0,-2.85)}%
\mov(2.,0){\Ellipse*(0.5)[4,1]}%
\mov(7.,0){\Ellipse*(0.5)[4,1]}%
\mov(7.75,3){\wavelin(-1.5,-2.85)}%
\mov(6.25,3){\wavelin(1.5,-2.85)}%
}
\vspace{5mm}
\caption{Feynman diagrams for proton polarizability correction
to the hydrogen HFS}
\end{figure}

The proton spin structure functions can be measured in the inelastic
scattering of polarized electrons on polarized protons. Recent improvements
in polarized lepton beams and targets have made it possible to make
increasingly accurate measurements of nucleon polarized structure
functions $g_{1,2}$ in experiments at SLAC, CERN and DESY [12-19]. The spin dependent
structure functions may be expressed in terms of virtual photon-absorption
cross-sections \cite{Close}:
\begin{equation}
g_1(\nu,Q^2)=\frac{m_p\cdot K}{8\pi^2\alpha(1+Q^2/\nu^2)}\left[\sigma_{1/2}
(\nu,Q^2)-\sigma_{3/2}(\nu,Q^2)+\frac{2\sqrt{Q^2}}{\nu}\sigma_{TL}(\nu,Q^2)\right]
\end{equation}
\begin{equation}
g_2(\nu,Q^2)=\frac{m_p\cdot K}{8\pi^2\alpha(1+Q^2/\nu^2)}\left[-\sigma_{1/2}
(\nu,Q^2)+\sigma_{3/2}(\nu,Q^2)+\frac{2\nu}{\sqrt{Q^2}}\sigma_{TL}(\nu,Q^2)\right]
\end{equation}
where $K=\nu-\frac{Q^2}{2m_p}$ is Hand kinematical flux factor for virtual
photons, $\sigma_{1/2}$, $\sigma_{3/2}$ are the virtual photoabsorption
transverse cross sections for total helicity between photon and nucleon
of 1/2 and 3/2 respectively, $\sigma_{TL}$ is the interference term
between the transverse and longitudinal photon-nucleon amplitudes. In this
work we calculate contribution $\Delta E_{HFS}^P$ on the basis of modern
experimental data on structure functions $g_{1,2}(\nu,Q^2)$ and theoretical
predictions on cross sections $\sigma_{1/2,3/2,TL}$.

To obtain correction (3) at the resonance region ($W^2\leq 4 GeV^2$) we use the
Breit-Wigner parameterization for the photoabsorption cross sections in
(12), (13), suggested in \cite{Walker,Arndt,Teis1,Teis2,Krusche,Bianchi,D}. There are many
baryon resonances that give contribution to photon absorption cross sections.
We take into account only five important resonances: $P_{33} (1232)$,
$S_{11} (1535)$, $D_{13} (1520)$, $P_{11} (1440)$, $F_{15} (1680)$.
Considering the one-pion decay channel of the resonances, the absorption
cross sections $\sigma_{1/2}$ and $\sigma_{3/2}$ may be written as follows
\cite{Teis2,Dong1}:
\begin{equation}
\sigma_{1/2,3/2}=\left(\frac{k_R}{k}\right)^2\frac{W^2\Gamma_\gamma\Gamma_{R
\rightarrow N\pi}}{(W^2-M_R^2)^2+W^2\Gamma_{tot}^2}\frac{4m_p}{M_R\Gamma_R}
|A_{1/2,3/2}|^2
\end{equation}
where $A_{1/2,3/2}$ are transverse electromagnetic helicity amplitudes,
\begin{equation}
\Gamma_\gamma=\Gamma_R\left(\frac{k}{k_R}\right)^{j_1}\left(\frac{k_R^2+X^2}
{k^2+X^2}\right)^{j_2},~~X=0.3~{\rm GeV}.
\end{equation}
The resonance parameters $\Gamma_R$, $M_R$, $j_1$, $j_2$, $\Gamma_{tot}$
were taken from \cite{RPP,Teis3}. In accordance with Refs. \cite{Teis1,Krusche,Teis3}
the parameterization of one-pion decay width is
\begin{equation}
\Gamma_{R\rightarrow N\pi}(q)=\Gamma_R\frac{M_R}{m_p}\left(\frac{q}{q_R}\right)^3
\left(\frac{q_R^2+C^2}{q^2+C^2}\right)^2,~~C=0.3~{\rm GeV}
\end{equation}
for the $P_{33}(1232)$ and
\begin{equation}
\Gamma_{R\rightarrow N\pi}(q)=\Gamma_R\left(\frac{q}{q_R}\right)^{2l+1}
\left(\frac{q_R^2+\delta^2}{q^2+\delta^2}\right)^{l+1},
\end{equation}
for $D_{13}(1520)$, $P_{11}(1440)$, $F_{15}(1680)$. l is the pion angular
momentum and $\delta^2=(M_R-m_p-m_\pi)^2+\Gamma_R^2/4$. Here k and q are
the photon and pion 3-momentum in the cms for a given center of mass energy
W, $k_R$ and $q_R$ are taken at the pole of the resonance. In the case
of $S_{11}(1535)$ we take into account $\pi N$ and $\eta N$ decay modes
\cite{Krusche,Teis3}:
\begin{equation}
\Gamma_{R\rightarrow\pi,\eta}=\frac{q_{\pi,\eta}}{q}b_{\pi,\eta}\Gamma_R
\frac{q_{\pi\eta}^2+C_{\pi,\eta}^2}{q^2+C_{\pi,\eta}^2},
\end{equation}
where $b_{\pi,\eta}$ are the $\pi$ ($\eta$) branching ratio.

The cross section $\sigma_{TL}$ is determined by expression similar to (14),
containing product $(S^\ast_{1/2}\cdot A_{1/2}+A_{1/2}^\ast S_{1/2})$ \cite{Abe1}.
The calculation of helicity amplitudes $A_{1/2}$, $A_{3/2}$ and longitudinal
amplitude $S_{1/2}$, as functions of $Q^2$, was done on the basis of
constituent quark model in \cite{Isgur,CL,Capstick,LBL,Warns}. In the real photon
limit $Q^2=0$ we take corresponding resonance amplitudes from \cite{RPP}.
For the $\Delta$ - isobar amplitudes $A_{1/2}(Q^2)$, $A_{3/2}(Q^2)$ we used
relations obtained in \cite{Carlson}. Helicity amplitudes of the other
resonances were taken from \cite{CL,Capstick,LBL,Warns}. We have
considered Roper resonance $P_{11}(1440)$ as ordinary $qqq$ state. As
it follows from predictions of the quark model, the helicity amplitudes,
which may be suppressed at $Q^2=0$, become dominant very rapidly with $Q^2$.
It may be seen on Figures 2-5, where we have shown also experimental data
of E143 collaboration at two fixed momentum transfer points:
$Q^2\approx 0.5 GeV^2$ and $Q^2\approx 1.2 GeV^2$. Our results for structure
function $g_1(\nu,Q^2)$ on Figures 2-3, which are in qualitative agreement
with \cite{Dong1} and experimental data,
show that Breit-Wigner five resonance parameterization of photon cross
sections and constituent quark model results give good description of proton
polarized structure functions at the resonance region.
The existing difference of this model for $g_{1,2}(\nu,Q^2)$ and experimental
data, which is particulary seen in Figure 3, demands further improvement
in the construction of spin dependent structure functions. This may be done
considering contributions of other baryonic resonances in the large W range:
$S_{31}(1620)$, $F_{37}(1950)$, $D_{33}(1700)$, $P_{13}(1720)$, $F_{35}(1905)$
and accounting different decay modes of such states \cite{Dong1}.
The sum rule of Gerasimov-Drell-Hern \cite{GDH}
\begin{equation}
-\frac{\kappa^2}{4m_p^2}=\frac{1}{8\pi^2\alpha}\int_{\nu_{th}}^\infty
\frac{d\nu}{\nu}[\sigma_{1/2}(\nu,0)-\sigma_{3/2}(\nu,0)].
\end{equation}
is valid with high accuracy \cite{Dong1}. The second part of (4) gives
especially large negative contribution to the correction $\delta_1^P$
in the range of small $Q^2$, where the contribution of $\Delta$ isobar
is dominant. With increasing $Q^2$ it's value falls
and total correction $\delta_1^P$ has positive sign.

\begin{figure}[htbp]\vspace*{0.0cm}
\epsfxsize=0.9\textwidth
\centerline{\psfig{figure=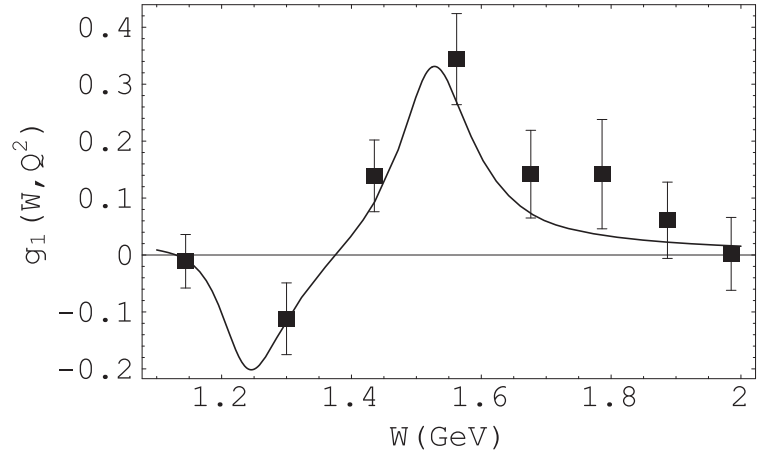,height=6.0cm,width=10.0cm}}
\caption{Proton structure function $g_1(W,Q^2)$ for $Q^2=0.5$ in
the resonance region.
Experimental points correspond to the paper [12]}
\end{figure}

\begin{figure}[htbp]\vspace*{0.0cm}
\epsfxsize=0.9\textwidth
\centerline{\psfig{figure=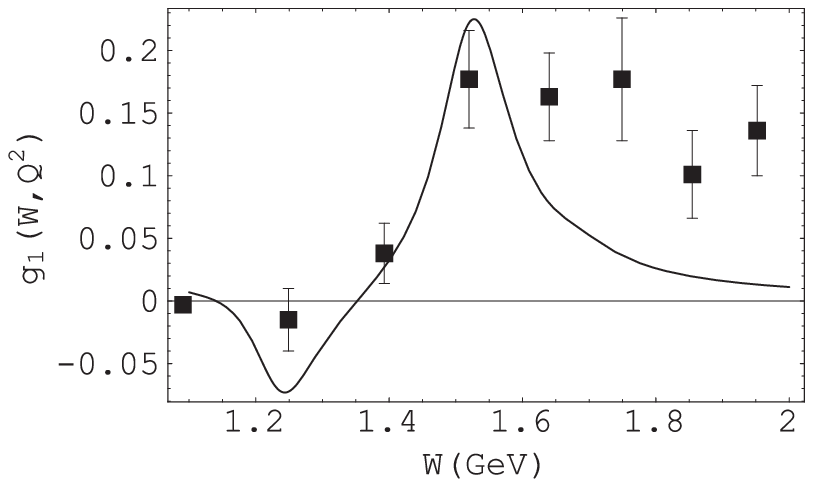,height=6.0cm,width=10.0cm}}
\caption{Proton structure function $g_1(W,Q^2)$ for $Q^2=1.2$ in
the resonance region.
Experimental points correspond to the paper [12]}
\end{figure}

\begin{figure}[htbp]\vspace*{0.0cm}
\epsfxsize=0.9\textwidth
\centerline{\psfig{figure=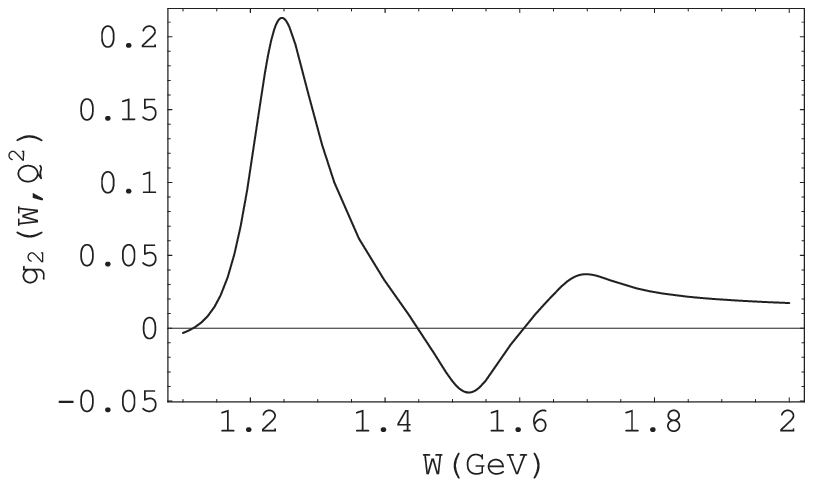,height=6.0cm,width=10.0cm}}
\caption{Proton structure function $g_2(W,Q^2)$ for $Q^2=0.5$
in the resonance region.}
\end{figure}

\begin{figure}[htbp]\vspace*{0.0cm}
\epsfxsize=0.9\textwidth
\centerline{\psfig{figure=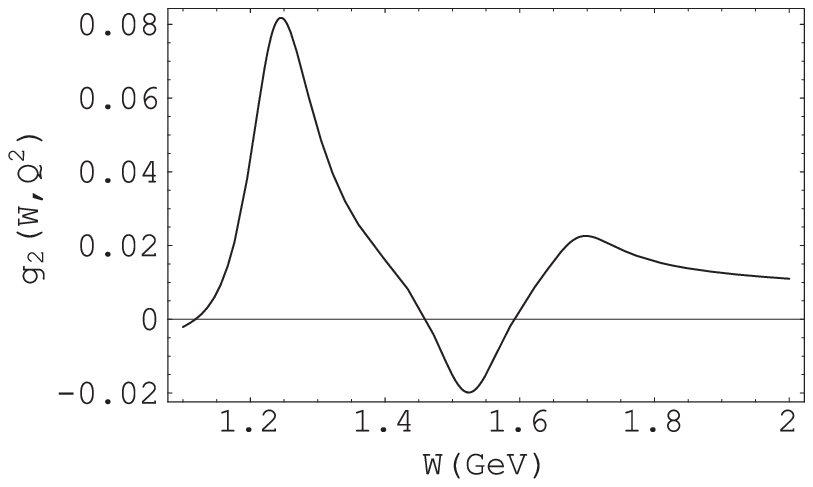,height=6.0cm,width=10.0cm}}
\caption{Proton structure function $g_2(W,Q^2)$ for $Q^2=1.2$
in the resonance region.}
\end{figure}

\begin{figure}[htbp]\vspace*{0.0cm}
\epsfxsize=0.9\textwidth
\centerline{\psfig{figure=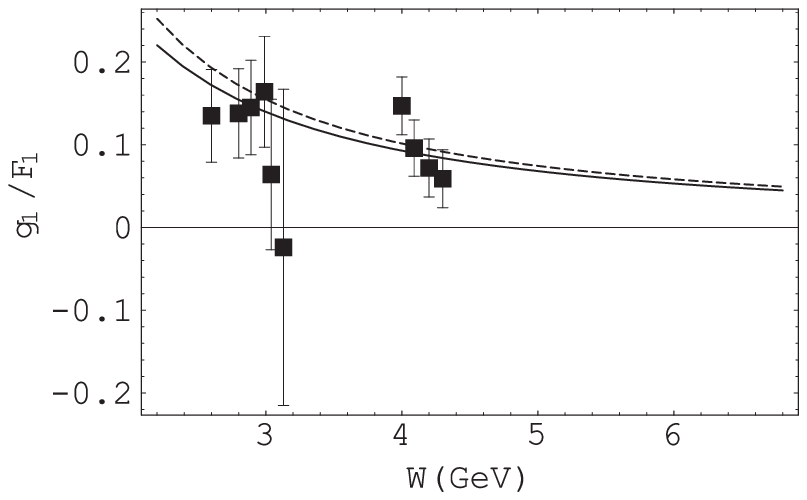,height=6.0cm,width=10.0cm}}
\caption{$g_1(W,Q^2)/F_1(W,Q^2)$ as a function of W for proton
at $Q^2=0.7$ in DIS region [12]. The dashed and solid curves are the
results of fits I and IV respectively.}
\end{figure}

\begin{figure}[htbp]\vspace*{0.0cm}
\epsfxsize=0.9\textwidth
\centerline{\psfig{figure=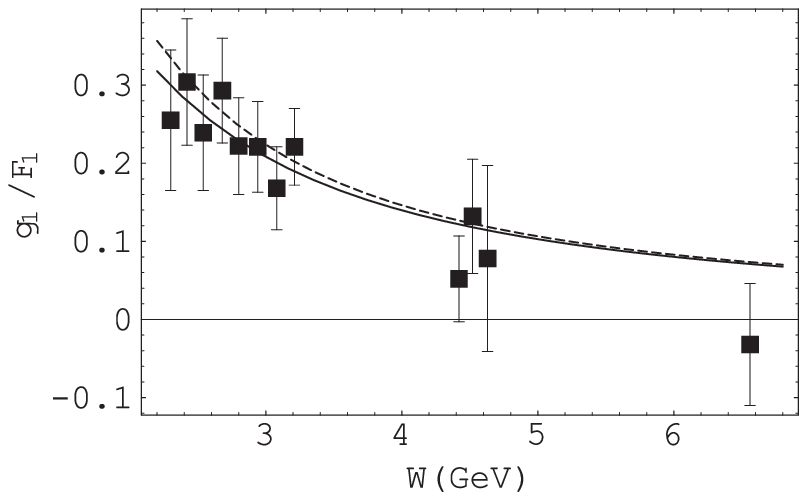,height=6.0cm,width=10.0cm}}
\caption{$g_1(W,Q^2)/F_1(W,Q^2)$ as a function of W for proton
at $Q^2=1.2$ in DIS region [12]. The dashed and solid curves are the
results of fits I and IV respectively.}
\end{figure}

Our calculation of the contribution $\Delta E_{HFS}^P$ in the DIS region
($W^2\geq 4 GeV^2$) is based on recent experimental data [12-19]. All of the data,
including the SMC data at $Q^2\leq 1 GeV^2$, were fit to the parameterization:
\begin{equation}
g_1(x,Q^2)=a_1 x^{a_2}(1+a_3x+a_4x^2)[1+a_5f(Q^2)]F_1(x,Q^2),
\end{equation}
where $x=Q^2/2m_p\nu$ is the Bjorken scaling variable, $F_1=W_1m_p$.
The coefficients of the fits and different models for the form of the
$Q^2$ dependence may be found in \cite{Abe1,E}. In Figures 6-7 the experimental
data and parameterization in the form (20) for the ratio $g_1/F_1$ are
presented at two different points $Q^2$. Numerical integration in (4) was
performed with the $f(Q^2)=-\ln Q^2$ (fit IV), corresponding to pQCD behavior.
We have extrapolated relation (20) to the region near $Q^2=0$.
Calculation of the second part of correction $\delta^P$ in (5) for
nonresonance region was performed by means of the Wandzura-Wilczek relation
between spin structure functions $g_1(x,Q^2)$ and $g_2(x,Q^2)$:
\begin{equation}
g_2(x,Q^2)=-g_1(x,Q^2)+\int_x^1 g_1(t,Q^2)\frac{dt}{t}.
\end{equation}

The values of contributions $\delta_1^P$, $\delta_2^P$ and total
contribution $\delta^P$, obtained after the numerical integration
in the resonance and nonresonance regions are as follows:
\begin{equation}
\delta_{1,res}^P=0.93~{\rm ppm},~~~\delta_{1,nonres}^P=0.86~{\rm ppm},~~\delta_1^P=1.79~{\rm ppm},
\end{equation}

\begin{equation}
\delta_{2,res}^P=-0.42~{\rm ppm},~~~\delta_{2,nonres}^P=-0.01~{\rm ppm},~~\delta_2^P=-0.43~{\rm ppm},
\end{equation}

\begin{equation}
\delta^P=\delta_1^P+\delta_2^P=1.4\pm 0.6~{\rm ppm},
\end{equation}
where the error, indicated in the expression (24), is determined by two
main factors, connected with the polarized structure functions:
uncertainty of the experimental data in the nonresonance
region and possible contribution of the other baryonic resonances
to the functions $g_{1,2}(\nu,Q^2)$. Estimation of the second error
was done by means of the integration results in (4), (5) for the different
intervals of $Q^2$, ${\rm W}$ and possible modification of the spin dependent
structure functions in the resonance region ${\rm W}\geq 1.5$ GeV
due to changing of the Breit-Wigner parametrization (14). First part of the error
in (24) is connected with statistical and systematical errors of the
experimental data from \cite{Abe1}.

The difference between the experimental value (1) and the
theoretical result $\Delta E_{HFS}^{th}$
without the proton polarizability contribution can be presented
in the form \cite{DTF,EGS,BY,KSG}:
\begin{equation}
\frac{\Delta E_{HFS}^{exp} - \Delta E_{HFS}^{th}}{E_F}=4.5 (1.1)~{\rm ppm}
\end{equation}
As was pointed out in \cite{DTF,EGS,BY}, the main sources of uncertainty
in this difference are the inaccuracy of the proton form factor
parameterization (dipole fit etc.) and the contradictory experimental data
on the proton radius. The proton polarizability correction $\delta^P$
calculated here gives the contribution (24) of the
proper sign and order of magnitude to the difference (25).
Further improvement of this calculation is connected just as the new
experimental and theoretical investigation of the internal structure
of the light quark baryons, new more
accurate measurements of the proton polarized structure functions,
so with the using QCD-based methods of the spin dependent structure
functions calculation \cite{Altarelli,Kumano}.
More detailed consideration of the structure functions
$g_{1,2}(\nu,Q^2)$ at the resonance region, taking into account contributions
of some other baryonic resonances and additional decay channels is also needed.
This work is in the progress.

We are grateful to M.I. Eides, S.G. Karshenboim, I.B. Khriplovich,
V.A. Petrun'kin, R.A. Sen'kov, for useful discussions.
The work was performed under the financial
support of the Russian Foundation for Fundamental Research
(grant 00-02-17771) and the Program "Universities of Russia - Fundamental
Researches" (grant 990192).

\end{document}